# Pseudo-Random Generator based on a Photonic Neuromorphic Physical Unclonable Function

Dimitris Dermanis, Panagiotis Rizomiliotis, Adonis Bogris, and Charis Mesaritakis

*Abstract*—In this work we provide numerical results concerning a silicon-on-insulator photonic neuromorphic circuit configured as a physical unclonable function. The proposed scheme is enhanced with the capability to be operated as an unconventional deterministic pseudo-random number generator, suitable for cryptographic applications that alleviates the need for key storage in non-volatile digital media. The proposed photonic neuromorphic scheme is able to offer NIST test compatible numbers with an extremely low false positive/negative probability below $10^{-14}$. The proposed scheme offers multi-functional capabilities due to the fact that it can be simultaneously used as an integrated photonic accelerator for machine-learning applications and as a hardware root of trust.

## I. INTRODUCTION

RANDOM numbers are the foundation of cryptography and cyber-security by offering cryptographic keys that, in turn, safeguard the anonymity and data integrity in both communication and computing. Random number generators can be classified into two basic categories based on their underlying principle of operation; namely true random number generators (TRNG) and pseudo-random number generators (PRNG). The first entails true stochastic processes while, PRNGs rely on a publicly known algorithm that generates a number sequence. Although the latter shares common statistical properties with TNRGs, it is deterministic. In real-world installments, PRNGs vastly dominate the landscape, due to the fact that they can offer higher random bit rate and do not require dedicated hardware. On the other hand, their algorithmic nature renders them vulnerable to a series of potential exploits [1].

Based on the above, there is a growing need for PRNGs that do not rely on software solutions, but their security feature stems from their physical properties. In this context, physical roots of trust [2], where randomness originates from the physical/complex features of an object or system can provide a solid foundation for building modern secure systems. Physical Unclonable Functions (PUFs) are the most well-known embodiments of the physical root of trust concept. They harvest their physical cloning resiliency from inevitable imperfections in their manufacturing process, whereas their PRNG function relies on their intrinsic complex physics. Typically, an input (referred to as a 'challenge') is launched into a PUF in order to generate a unique output (the 'response'). This response is the deterministic outcome of a highly complex physical function, which is distinct for each device and is computationally irreversible. These challenge-response pairs (CRPs) are used as a type of hardware based fingerprints (weak PUF) or can be considered a physics driven PRNGs (strong PUF) [3].

The first demonstration of a PUF relied on bulk optics [4] and was able to produce unique responses by illuminating a token (diffuser) using a laser beam. Despite the merits of this first primitive, key drawbacks arose such as the limited number of CRPs [5], hindering PRNG operation, and the fact that optical PUFs rely on bulky components and are difficult to co-integrate with electronics. Silicon cast PUFs based on CMOS electronics, followed, relying on unpredictable, fabrication related variations, in features such as the length of electrical delay lines, gate-voltage in transistors and SRAM's initial state [6], [7]. Electronic PUFs have been readily deployed as authentication tokens and as PRNGs, but they rely on simpler physical mechanisms compared to optics, thus are rendered vulnerable to a plethora of attacks based on modelling, machine learning and side-channel [8], [9] [10], [11],[12].

In recent years, a new generation of more exotic PUF types have been proposed relying on new platforms or offering new designs to existing ones. Among them there are approaches that leverage the molecular properties of the materials used [13], [14] and exploit the randomness of surfaces or volumes [15], [16], [17]. Memristor PUFs [18], [19], show considerable potential, while photonic alternatives based on silicon on insulator [20], [21], [22], although not yet mature, are also gaining industrial traction, offering increased machine learning attack resiliency [23].

This work was supported by EU Horizon project PROMETHEUS under Grant 101070195. *Corresponding author: cmesar@aegean.gr*

D. Dermanis and C. Mesaritakis are with the department of Information and Communication Systems Engineering, University of the Aegean – Greece. A. Bogris is with the department of Informatics and Computer Engineering, University of West Attica – Greece and P. Rizomiliotis is with the department of Informatics and Telematics, Harokopio University – Greece.

In our previous work [24], we have introduced a new concept of a neuromorphic photonic PUF (nPUF) and demonstrated its physical unclonability characteristics relying on standard silicon-on-insulator (SOI) manufacturing yield. The proposed nPUF offered unique advantages in terms of emulation and machine learning attack resiliency, due to its neuromorphic driven temporal dynamics. In this work, we build upon our initial design and offer numerical simulations confirming that the nPUF concept can be configured to operate also as a secure clone resilient PRNG [2]. Towards this direction we investigated hardware related parameters and CRP post-processing techniques so as to assess CRPs consistency and randomness. With an average inter-challenge hamming distance of 0.46, a beyond state-of-the-art equal error rate (EER) of $10^{-14}$ and NIST statistical tests compatibility, we argue that the proposed system can be used as a multi-application platform addressing both ML tasks [25], [26] and simultaneously harden the security features of edge-devices.

## II. PRINCIPLE OF OPERATION

In order to clarify the basic operation of a nPUF we will first present, in an implementation agnostic fashion, the principle of operation. In particular, the core of the nPUF resides in the existence of a dynamical recurrent neural network (RNN), such as reservoir computing (RC) [27]. The RC in our case, is well suited, due to the fact that it preserves the complex dynamical response of the RNNs, while adopting a hardware friendly – "random" approach to its synaptic connections. The RC's hidden "random" layer is assumed to be implemented in a hardware platform, where fabrication imperfections dictate the existence of non-controllable parameter deviations (in weights, bias etc.). This layer is followed by a digital trainable readout layer, where the weights can be controlled with a high bit-accuracy. The challenge in our case, consists of two parts; an input time-trace and a target time-trace that can be drawn from a random distribution or can be inter-connected through a complex mathematical formula (e.g. NARMA [28], Mackey-Glass [29] etc.). The input time-trace is injected to the hardware-based RC and its output states are detected (e.g. from a photodiode), are digitized through an analogue-to-digital converter (ADC) with 8-bit resolution and are fed to the digital readout. This layer is subsequently trained so as the input trace fits to the target trace. Based on this operation all weight imperfections at the RC's hidden layer will manifest to the trainable digital weights, following a dynamical, non-linear relation. In particular, as shown in Fig. 1, by following typical RC formalism we derive the following relations: if the timeseries pair is $\{X_{in}, Y_{out}\}$ then $X_{in}$ is injected to the RC.

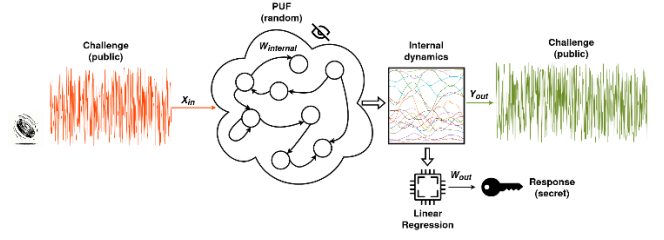

**Fig. 1.** Schematic of the neuromorphic PUF concept. The $X_{in}$ part (public) of the challenge drives the network`s internal states, also affected by the $W_{internal}$ that is kept secret. Then, the network is trained using these internal states and the $Y_{out}$ part (public) of the challenge. The training weights $W_{out}$ are the PUF`s response.

The time-varying relationship between the input $X_{in}$ and each RC's state ($S$) is described in (1), where $f$ corresponds to the RC's nonlinear activation function, $n$ represents additive white Gaussian noise and $k_i$ is the delay associated with each one of the $N$ interconnected nodes $i$. This is a general formalism of the concept that adapts to the platform used for the implementation of the RC scheme.

$$S[t] = f(W_{int} \cdot S[t - k_i] + X_{in}[t] + n_i), i = 1 \ldots N \quad (1)$$

It is clear that $S$ depends on both the internal structure of the weight matrix $W_{int}$ and the input $X_{in}$. following this step, the readout layer is trained e.g. through a ridge regression, thus solving $Y_{out} = W_{out} \cdot S \Rightarrow W_{out} = Y_{out} \cdot S^{-1}$ results in the output digital weights $W_{out}$. Therefore, the dynamic dependence of $S$ on $W_{int}$ and $X_{in}$ is subsequently conveyed to $W_{out}$, encapsulating both the RC's inherent structure and the challenge problem's features. In [24] we assumed that the same challenge was used in all cases, and we generated multiple nPUF instances. Through this approach the probability of cloning was computed and nPUFs were considered as a single CRP authentication token. Although physical unclonability and emulation resiliency was validated, the real merits of such an optical nPUF would shine if PRNG operation is achieved, thus offering a high-speed, integrated and unclonable key generation and storage solution. Therefore, here we assume a single nPUF hardware instance and we evaluate the variation at $W_{out}$ (response) versus $\{X_{in}, Y_{out}\}$ pairs (challenge).

Without loss of generality, as CRP mechanism we utilize a Nonlinear Autoregressive Moving Average (NARMA) timeseries [28] described by (2) where $a_1$, $a_2$, b, c are constant parameters, and $X_{in}$ is drawn from a uniform distribution ranging from 0 to 1. For this work, the values $a_1 = 0.3$, $a_2 = 0.05$, b = 1.5, c = 0.1 were used.

$$Y_{out}[t + 1] = a_1 Y_{out}[t] + a_2 y[t] \left( \sum_{i=0}^{m-1} Y_{out}[t - 1] \right) + b X_{in}[t - (m - 1)] X_{in}[t] + c \quad (2)$$

## III. METRICS

After defining the principle of operation, the next step is to define generic metrics that can assess nPUF's performance and security. In this context, several metrics that collectively ascertain its security, reliability, and practical applicability are considered. The first important metric used in this work for evaluation is reproducibility. Reproducibility (or robustness) measures the capability of the same PUF to produce an 'identical' response to the same challenge over time, considering varying environmental conditions. This metric is quantified by measuring the hamming distance among repeated acquisitions of the PUF's digitized responses under identical input. The second critical metric used here is identifiability. Identifiability measures the ability of a PUF to produce different responses as a result of different challenges. To quantify this property of the proposed PUF, a big collection of challenges is injected to the PUF and the hamming distances of the responses is measured.

The combined evaluation of identifiability together with reproducibility can be perceived using the equal error rate (EER), which shows the point where the false acceptance rate equals the false rejection rate. It will be used here to determine the balance point of security and usability of our concept. While randomness can occur through many sources including noise, in PUF concept the hardware should work as a "deterministic" PRNG and thus be able to reproduce the response tied to a specific challenge. Finally, the randomness of the PUF's robust responses is tested by using the well-established NIST statistical test suite [30].

## IV. PHOTONIC IMPLEMENTATION

As mentioned above, a hardware manifestation of an RC is needed at the core of the nPUF to provide its security features. Similar to [24] we employ a recurrent optical spectrum slicing (ROSS) neuromorphic scheme, depicted in Fig. 2. Briefly it consists of a single waveguide loop, generated through two 3dB couplers, whereas inside the loop there are one or more bandpass or bandstop filters. In this configuration the filters consist of micro-ring resonators (MRR) in an add/drop configuration (inset of Fig. 2).

The $X_{in}$ is assumed to amplitude modulate an optical carrier at a wavelength of 1556nm with a mean power of 10 dBm. The optical signal is injected through a 1×4 splitter to four ROSS nodes, whereas each node includes six add/drop MRRs in series. Each MRR's resonant frequency is positioned so as to target a different spectral regime of the incoming signal with an inter-MRR detuning of 1 GHz whereas each MRR bandwidth is in the order of 1 GHz (inset of Fig. 2). The number of nodes was chosen so as to partially cover the full extent of the signal's bandwidth. The drop ports of each MRR are sent to photodiodes (PD) with a bandwidth of 40 GHz matching the time-series modulation rate (40 Gsymbol/sec), while PDs are considered to be subject to both thermal and shot-noise.

The analogue outputs of the PDs are fed to an equal number of analogue-to-digital converters (ADC), with a sampling rate of 40 Gsample/sec, resulting to one sample per symbol. The ADC bit-accuracy is considered tunable and ranges ($m_{bit}$) from 16 bit to 1 bit, thus allowing varying quantization noise. Although the state-of-the-art ADCs resolution at 40 Gsample/sec is 10 bit, we expanded our research in order to give some insights for potential lower baudrate implementations. The digitized data from the ADCs that correspond to the RC's states ($S$) are sent to a computer, where a lightweight linear regression model is implemented. In order to address the NARMA-10 task, we opted for storing 11 outputs for each symbol, since the current value depends on the 10 previous values. In this context, for a $N$-filter scheme the values used for the regression model are $N\times 11 + 1$ per symbol, with the extra one being the direct input to output connection. As clarified in section II, $S$, $X_{in}$ and $Y_{out}$ are fed to the linear regression model so as to extract digital $W_{out}$, which in our case is the PUF's response.

In this nPUF manifestation, the main uncontrollable physical parameter relies on standard SOI specifications regarding the waveguide roughness [31], [32]. In this context, key waveguide components such as loop, MRRs, etc. are assumed to exhibit a random refractive index variation ($\Delta n_{eff}$) that in turn corresponds to a random inter-chip phase bias, among different SOI chips. According to [30-31] the inter-chip variation in waveguide roughness can result to a $\Delta n_{eff}$ following a uniform distribution, with values ranging from -0.015 to 0.015. Taking into consideration that the components used in this architecture are resonant devices, these $\Delta n_{eff}$ lead to unpredictable resonant frequency shifts and Q-factor variations. The ROSS and the underlying physical uncertainties were numerically simulated in [24], where these features were used to confirm the unclonability of multiple nPUFs. All the associated parameters of the system are included in Table 1.

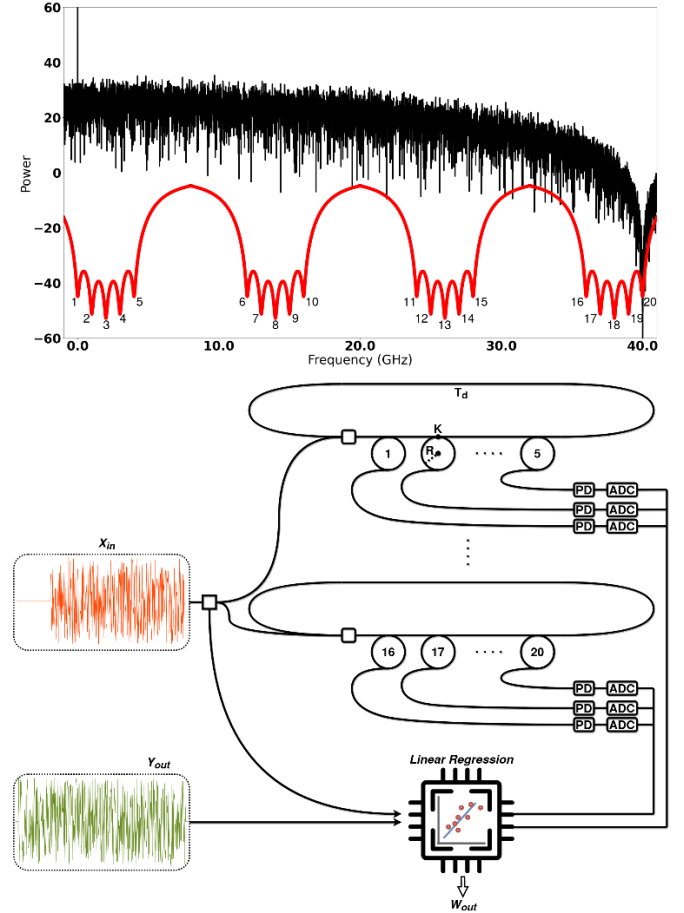

**Fig. 2.** Photonic implementation using 20 MRRs arranged in 4 banks of 5 MRR each that filter the input timeseries. Each MRR is centered on a specific part of the input spectrum (inset – black), denoted by the corresponding number on the combined transfert function of the MMRs (inset – red).

Operation wise, here we assume a single nPUF and we generate multiple CRPs to test PRNG capabilities. Towards this direction,

the $X_{in}$ component of each CRP was drawn from a uniform distribution [-1, 1], generated through a pseudo-random number generator (PCG64), while the $Y_{out}$ component of the CRP was derived using equation (3). To accumulate sufficient data for assessing the randomness of the responses, we repeated the simulation process $5\times10^5$ times. In the ridge regression training procedure, 2000 NARMA samples were sufficient to reproduce the sequence with a mean normalized mean square error (NMSE) = 0.023 [24]. The generated weights ($W_{out}$) for each NARMA sequence were post-processed as follows: The occurring weights follow a Gaussian distribution $X = N(\mu=0, \sigma=3)$, that is initially normalized to become a normal distribution $X' = N(\mu=0, \sigma=1)$. By computing the cumulative distribution function (CDF) $\Phi_X$ and then finding the $\Phi_X(w_i)$ for each weight $w_i$, we end up with a uniform distribution of numbers in [0, 1], according to inverse transformation sampling (or inversion sampling) [33]. It is crucial to note that for the calibration phase, the CDF was computed not from a single CRP, but rather from an ensemble of such pairs. In our simulations, this ensemble consisted of $10^3$ pairs. However, this size can be adjusted downward, provided that a sufficient number of weights are utilized to ensure a densely populated distribution. The final step consists of quantizing the analogue weights to $n_{bit}$, which in turn is related to the number of bins ($b$) used to quantize $\Phi_X$, ($n_{bit}=\log_2 b$). The $n_{bit}$ number essentially controls the ability of the system to reproduce the generated random numbers, but also the randomness of the bits generated. Higher $n_{bit}$ values will dissect the space into more bins, thus making it easier for a value to be found in neighboring bins and produce different binary results, while lower values will make the system resilient to noise because less bins results to increased probability for a value to be constantly at the same bin.

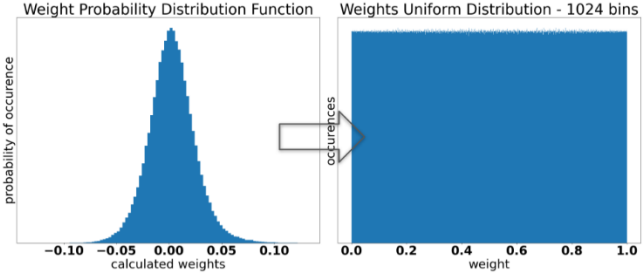

**Fig. 3.** The weights calculated (left) follow a normal distribution. They are processed to a uniform distribution (right) using inverse transformation sampling.

| PARAMETER | VALUE | DEVIATION |
|---|---|---|
| κ (coupling coefficient) | 0.25 | constant |
| $F_{str}$ (feedback strength) | 0.9 | constant |
| $T_d$ (loop delay) | 25ps | constant |
| $T_{MRR}$ (inter-MRR delay) | 2.5ps | constant |
| $f_{MRR}$ (inter-MRR frequency separation) | 1GHz | N (μ=$f_{ideal}$, σ=0.1) |
| $C_{MRR}$ (inter-MRR connection strength) | 0.95 | N (μ=0.97, σ=0.1) |
| $n_{eff}$ (waveguide) | 3.4 | U (-0.015, 0.015) |
| a (propagation losses) | $10m^{-1}$ | constant |
| R (ring radius) | 55μm | constant |

**Table 1:** Photonic RC simulation parameters

## V. PHOTONIC nPUF AS STANDALONE PRNG

A key aspect of PUF based PRNGs is deterministic operation; meaning that the output numbers should not be solely the product of a stochastic process (e.g. noise) but they should be reproduced on demand if the correct nPUF-CRP combo is used. This feature is critical for "unconventional" key storage, where keys are stored in the physical structure of the system, thus not in typical cyber-vulnerable digital media. Toward this end, $W_{out}$ perceived randomness, although it depends on the system's noise (thermal, shot-noise, laser noise etc.), it should be mainly governed by $W_{int}$-$X_{in}$ changes ($X_{in}$ used here). In addition, quantization noise introduced at the ADCs ($m$-bits) and by the number of bits ($n_{bit}$) used at binarization can affect nPUF's randomness. Taken into consideration that if optical power and system losses are assumed fixed, then these parameters can be used so as to regulate the impact of noise to the nPUF's responses.

In order to visualize the above-mentioned interplay between stochastic and deterministic mechanisms, in Fig. 4 we present three heatmaps with respect to $n_{bit}$ and $m_{bit}$. In Fig. 4a the intra hamming distance between outputs extracted by using the same CRP multiple times is presented, thus it is a measure of the system's noise to the output's randomness. It can be seen that by increasing $m_{bit}$ we directly reduce quantization noise, thus pushing the system towards low hamming values. On the contrary, by increasing the number of bins ($n_{bit}$) hamming distance increases. This effect can be attributed to the fact that if there are quantization noise induced variations at the RC's states (low ADC bit-accuracy), these in turn are translated to significant $W_{out}$ variations. On the other hand, after ridge regression by increasing the resolution that one binarizes $W_{out}$ allows small differences to be projected to large hamming distances. In Fig. 4a, the green curve demotes the area above which, the intra-hamming distance is low enough (<0.25) for PRNG, resulting to $n_{bit}$<5.

Complementary in Fig. 4b the inter-hamming distance is computed; meaning the hamming distance among outputs of different CRPs (as described in section IV). It can be seen that the reverse trend is observed: low $m_{bit}$ ($m_{bit}$<9) ensures a hamming close to the theoretical maximum of 0.5. Again, the green curve denotes the area below which the hamming distance is high enough for PRNG.

In Fig. 4c the EER is computed. In particular, an optimal EER is as low as possible, indicating a substantially low probability that the same challenge used twice will generate different responses and that two distinct challenges will yield responses perceived as identical. Consequently, one has to examine both these plots in order to determine the optimal parameters of a functional PUF device. In Fig. 4c we can clearly observe two distinct regions with a good EER score (<$10^{-12}$). Within these two regions the PUF demonstrates both reproducibility (low intra-challenge hamming distance) and identifiability (high inter-challenge hamming distance). If these regions are examined in combination with Fig. 4a-b (green curves of Fig. 4c) it can be seen that the upper region does not fulfill the unpredictability property (inter-hamming close to 0.5). In this context, by combining the green-curve defined areas of both Fig. 4a-b we can define a parameter space where the system offers robust yet unpredictable performance. The highest EER score is <$10^{-20}$, while in Fig. 4a-b defined zone a state-of-the art EER of $10^{-12}$ can be achieved.

The behavior exhibited in these two regions can be explained by considering the influence of certain parameters on the process of generating responses. Increasing the number of bits utilized for each weight representation magnifies the impact of small perturbations, potentially caused by noise, resulting in significant variations in the responses, rendering the system more unstable and consequently, more unpredictable. While this heightened unpredictability initially contributes positively to the system's performance, when surpassing a threshold point, the instability becomes detrimental, leading to a deterioration in the EER.

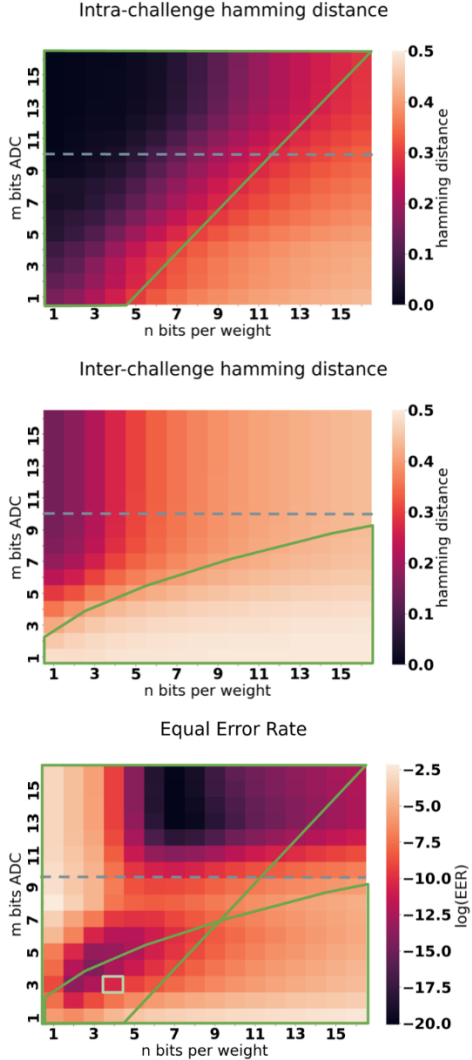

**Fig. 4.** (a) heatmap of the intra-challenge (intra-reservoir and intra-challenge terms both describe the repeated use of one reservoir instance and one challenge) hamming distance. (b) heatmap of the inter-challenge (responses given for different challenges) hamming distance. (c) heatmap of the equal error rate. The dashed gray line denotes the resolution limit for a state-of-the-art ADC at 40 Gsa/s.

From Fig. 4 it can be derived a set of parameters that simultaneously offer robust operation and uncorrelated responses. A typical case is when $m_{bit}$=3 and $n_{bit}$=4 bits (see Fig. 4 - green square). As mentioned, in section IV. Photonic Implementation, for a 24 MRR setup there will be 265 (24×11 + 1) training weights and as a result 1060 bits response for each CRP (265 weights × 4 bits / weight = 1060 bits). For this case in Fig. 5 we present the distribution of the different hamming distances. The blue histogram corresponds to the intra-hamming distance (single PUF, CRP subject to noise) it was found to be 0.22 ± 0.02. Next, the red histogram is the inter-challenge response hamming distance (red) of 0.46 ± 0.02. It can be seen that the two distributions do not visually overlap thus minimizing EER, where here is computed to be $10^{-10}$; sufficiently low to indicate a high level of security, according to [2]. Finally, we included in the assessment the uniqueness metric, a topic we explored in our prior work [24]. In this context, we simulated different PUF instances using a single CRP. The resulting inter-PUF response hamming distance (green) was 0.46 ± 0.02, reflecting the microscopic variations inherent to the hardware. Interestingly for the proposed nPUF it can be seen that variations in the CRP offer equally diverse responses as changing the hardware itself, thus the security analysis presented in [24] remains valid even for the case of a single device acting as PRNG.

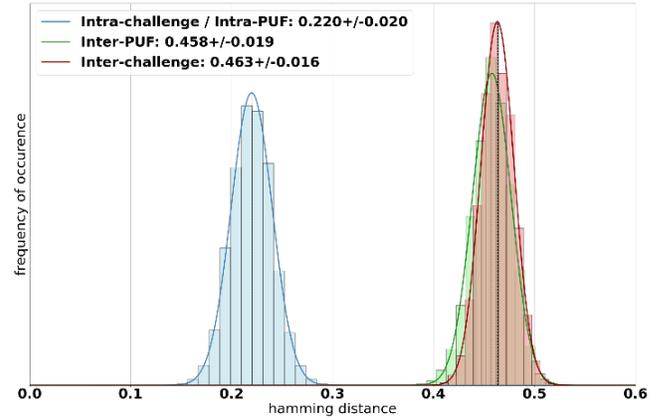

**Fig. 5.** Histograms of intra-challenge / intra-PUF (blue) hamming distances, inter-PUF (green) hamming distances and inter-challenge hamming distances (red), produced using all 5×$10^5$ simulation results.

A significant discussion is the number of components (MRRs) that the nPUF utilizes so as to offer its security features. First, the proposed neuromorphic module should be able to tackle the task that is associated with the CRP process, here NARMA. Given that a general rule of thumb is that MRRs' combined responses should be able to cover the input signal's bandwidth, it can be concluded that the number of MRRs is dictated by the modulation speed of the incoming data; higher rates dictate for more nodes. In our case, for a 40 Gbaud rate four loops with six MRRs are sufficient. Furthermore, it is important to note that varying the number of MRRs within each loop directly influences the total count of weights generated to solve the problem (output weights) and thus the final length of the binary sequence produced. In addition, the number of loops and MRRs affect emulation attack resiliency and cloning as shown in [24]. Here we will focus on how modifying the number of MRRs will affect robustness and correlation among responses so as to link chip size with PRNG capabilities.

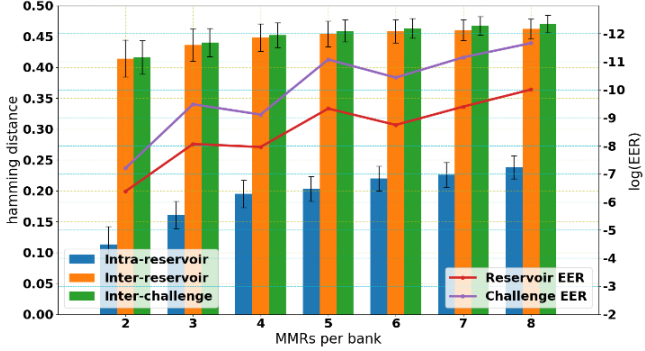

**Fig. 6.** EER and hamming distances against the number of MRRs per bank. The bars refer to hamming distances and the lines to EER.

In Fig. 6 we can see that as the number of MRRs per loop increases, an impact on stability is observed. Specifically, there is an improvement in both inter-PUF and inter-challenge hamming distances, with values converging closer to the ideal 0.5, which is indicative of lower correlation among responses. However, this comes at a cost, as the intra-PUF hamming distance also increases. The origin of this effect is the reduced signal-to-noise ratio at the PDs, due to the use of a constant input power (10 dBm) and the increase propagation and filtering losses associated with the use of more MRRs. Interestingly, the EER (solid lines) gradually increases with the increase of the MRR number. This can be related to the decrease of standard deviation bars in the hamming distances; meaning that although noise increases the mean value of the intra-hamming distribution, the increased number of weights that are injected to the linear regression reduce the impact of each individual weight, thus reduces the standard deviation. Therefore, an increased number of MRRs beyond 5 per loop, renders the scheme more unclonable [24] and at the same time reduces the correlation among responses and reduces the EER as shown in this work.

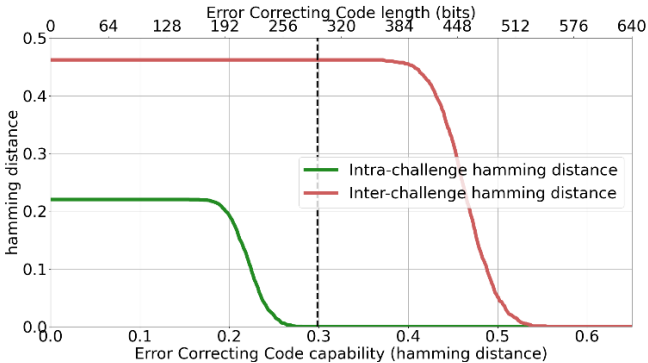

**Fig. 7.** Hamming distance correction for intra-challenge and inter-challenge simulations. The intra-challenge distance falls to zero well before the inter-challenge hamming distance is affected.

V. PHOTONIC nPUF PRNG FOR ENCRYPTION APPLICATIONS

The above analysis confirms the use of the proposed nPUF for authentication purposes. In this case, it is only needed to establish a decision threshold at the point between the intra-class and inter-class distribution curves, ensuring that the probabilities of False Acceptance (FA) and False Rejection (FR) are minimized. Setting the decision threshold at the EER provides a balanced trade-off, optimizing both security and functionality and allows standalone operation of the nPUF. On the other hand, when CRPs are utilized for key generation/storage that will be used by higher level encryption applications, it is imperative for the binary response to exhibit not marginal but zero-bit flips across repetitive nPUF interrogation. To achieve this consistency and rectify any bit flips the implementation of Error Correction Codes (ECC) is essential.

In order to demonstrate the feasibility of this approach we utilize a lightweight Bose–Chaudhuri–Hocquenghem error correcting code (BCH) so as to post-process nPUF's responses. Considering the initial binary response as the original key and the subsequent responses (using the same challenge) as authentication attempts, we evaluated the ability of the system to function properly accepting all responses generated by the original challenge (minimal FR) and rejecting all responses generated by other challenges (minimal FA). When a challenge is used for the first time, the ECC redundant bits are stored in public, as knowledge of these bits does not reveal information about the key. Thereafter, each time the challenge is used, these redundant bits are used to correct the re-produced key and obtain the original. In Fig. 7 we demonstrate the feasibility of correcting all potential bit flips induced by noise, without affecting the bit sequences generated by different challenges. Using an increasing number of ECC bits each time we tried to correct all the keys produced by the same challenge (intra-challenge) and all the keys produced be different challenges (inter-challenge). The upper x-axis demonstrates the extra bits needed to correct the 1060 bit key produced by the 24 MRR nPUF. It is clear that a distinct margin exists between the two distributions (300 < ECC bits < 380), and the intra-challenge hamming distance approaches zero significantly before the correction of any inter-challenge sequences occurs.

The final step in order to confirm randomness of the responses, inter-hamming distance is not enough, therefore, the NIST statistical test suite is employed. A total number of $5 \times 10^5$ binary responses from the PUF were concatenated allowing us to create a large enough dataset for running most of the tests using a confidence level of $\alpha=0.01$, as shown in Table 2. There are some tests (marked with an asterisk) that require a very extensive binary sequence. In order to run those tests as well, we had to extend the dataset accordingly. To do so we appended a random permutation of our dataset to the tail of the dataset. Given that our binary sequence is random, any permutation of it will also be random and this process does not add any randomness to the result [34]. So, the extended binary sequence consisting of the original sequence plus a random permutation of it was used to run the rest of the tests, maintaining the confidence level.

| P-VALUE | PROPORTION | STATISTICAL TEST | PASSED |
|---|---|---|---|
| 0.02750 | 990/1000 | Frequency | 1/1 |
| 0.20012 | 991/1000 | BlockFrequency | 1/1 |
| 0.16081 | 988/1000 | CumulativeSums | 2/2 |
| 0.99616 | 992/1000 | Runs | 1/1 |
| 0.80557 | 991/1000 | LongestRun | 1/1 |
| 0.51611 | 986/1000 | Rank | 1/1 |
| 0.09892 | 980/1000 | FFT | 1/1 |

| | | | | |
|---|---|---|---|---|
| 0.14950 | 983/1000 | NonOverlappingTemplate | 148/148 | |
| 0.05431 | 987/1000 | OverlappingTemplate | 1/1 | * |
| 0.10879 | 986/1000 | Universal | 1/1 | * |
| 0.75384 | 994/1000 | ApproximateEntropy | 1/1 | |
| 0.26249 | 561/565 | RandomExcursions | 8/8 | * |
| 0.52160 | 558/565 | RandomExcursionsVariant | 18/18 | * |
| 0.85800 | 993/1000 | Serial | 2/2 | |
| 0.96187 | 986/1000 | LinearComplexity | 1/1 | * |

**Table 2:** NIST statistical test results

Table 2 shows that all the tests are successful, thereby affirming the PUF's randomness capabilities.

## X. CONCLUSION

In this work, we demonstrate that the neuromorphic PUF we have introduced in [24] functions effectively as both a weak and a strong PUF, with randomness property verified by the NIST tests. This enables its application as an alternative to non-volatile key storage, facilitating the on-the-fly generation of cryptographically secure keys. Quantizing the outputs of the optical reservoir results in reduced deviation of mean values, as the quantization process mitigates some of the noise at the photodetector, up to a certain threshold. We have examined various configurations with different numbers of MRRs and determined the functional technical characteristics of such a system. Furthermore, we proposed a hybrid analog-digital system that combines the advantages of both worlds. It seamlessly integrates with modern digital computers and efficiently processes analog inputs at high-speed using less energy. Implemented with silicon photonics, the PUF is compact, enabling integration with various ecosystems, including IoT. This operational paradigm provides a twofold advantage: it leverages the lightspeed processing of neuromorphic computing, while utilizing the inherent physical randomness of the manufacturing process, ensuring unique identification and enhanced security for each device.

## ACKNOWLEDGMENT

This work has received funding from the EU Horizon Europe Program PROMETHEUS under grant agreement 101070195.

## REFERENCES

[1] J. Kelsey, B. Schneier, D. Wagner, and C. Hall, "Cryptanalytic Attacks on Pseudorandom Number Generators," in *Fast Software Encryption*, vol. 1372, S. Vaudenay, Ed., in Lecture Notes in Computer Science, vol. 1372. , Berlin, Heidelberg: Springer Berlin Heidelberg, 1998, pp. 168–188. doi: 10.1007/3-540-69710-1_12.
[2] R. Maes, *Physically Unclonable Functions: Constructions, Properties and Applications*. Berlin, Heidelberg: Springer Berlin Heidelberg, 2013. doi: 10.1007/978-3-642-41395-7.
[3] C. Herder, M.-D. Yu, F. Koushanfar, and S. Devadas, "Physical Unclonable Functions and Applications: A Tutorial," *Proc. IEEE*, vol. 102, no. 8, pp. 1126–1141, Aug. 2014, doi: 10.1109/JPROC.2014.2320516.
[4] R. Pappu, B. Recht, J. Taylor, and N. Gershenfeld, "Physical One-Way Functions," *Science*, vol. 297, no. 5589, pp. 2026–2030, Sep. 2002, doi: 10.1126/science.1074376.
[5] U. Rührmair *et al.*, "Optical PUFs Reloaded." 2013. [Online]. Available: https://eprint.iacr.org/2013/215
[6] B. Gassend, D. Clarke, M. Van Dijk, and S. Devadas, "Silicon physical random functions," in *Proceedings of the 9th ACM conference on Computer and communications security*, Washington, DC USA: ACM, Nov. 2002, pp. 148–160. doi: 10.1145/586110.586132.
[7] J. Guajardo, S. S. Kumar, G.-J. Schrijen, and P. Tuyls, "FPGA Intrinsic PUFs and Their Use for IP Protection," in *Cryptographic Hardware and Embedded Systems - CHES 2007*, vol. 4727, P. Paillier and I. Verbauwhede, Eds., in Lecture Notes in Computer Science, vol. 4727. , Berlin, Heidelberg: Springer Berlin Heidelberg, 2007, pp. 63–80. doi: 10.1007/978-3-540-74735-2_5.
[8] X. Xu and W. Burleson, "Hybrid side-channel/machine-learning attacks on PUFs: A new threat?," in *Design, Automation & Test in Europe Conference & Exhibition (DATE), 2014*, Dresden, Germany: IEEE Conference Publications, 2014, pp. 1–6. doi: 10.7873/DATE.2014.362.
[9] B. M. S. Bahar Talukder, F. Ferdaus, and M. T. Rahman, "Memory-Based PUFs are Vulnerable as Well: A Non-Invasive Attack Against SRAM PUFs," *IEEE Trans. Inf. Forensics Secur.*, vol. 16, pp. 4035–4049, 2021, doi: 10.1109/TIFS.2021.3101045.
[10] U. Ruhrmair *et al.*, "PUF Modeling Attacks on Simulated and Silicon Data," *IEEE Trans. Inf. Forensics Secur.*, vol. 8, no. 11, pp. 1876–1891, Nov. 2013, doi: 10.1109/TIFS.2013.2279798.
[11] M. Cortez, A. Dargar, S. Hamdioui, and G.-J. Schrijen, "Modeling SRAM start-up behavior for Physical Unclonable Functions," in *2012 IEEE International Symposium on Defect and Fault Tolerance in VLSI and Nanotechnology Systems (DFT)*, Austin, TX, USA: IEEE, Oct. 2012, pp. 1–6. doi: 10.1109/DFT.2012.6378190.
[12] Y. Gao, S. F. Al-Sarawi, and D. Abbott, "Physical unclonable functions," *Nat. Electron.*, vol. 3, no. 2, pp. 81–91, Feb. 2020, doi: 10.1038/s41928-020-0372-5.
[13] A. F. Smith, P. Patton, and S. E. Skrabalak, "Plasmonic Nanoparticles as a Physically Unclonable Function for Responsive Anti-Counterfeit Nanofingerprints," *Adv. Funct. Mater.*, vol. 26, no. 9, pp. 1315–1321, Mar. 2016, doi: 10.1002/adfm.201503989.
[14] N. Kayaci, R. Ozdemir, M. Kalay, N. B. Kiremitler, H. Usta, and M. S. Onses, "Organic Light-Emitting Physically Unclonable Functions," *Adv. Funct. Mater.*, vol. 32, no. 14, p. 2108675, Apr. 2022, doi: 10.1002/adfm.202108675.
[15] M. S. Kim, G. J. Lee, J. W. Leem, S. Choi, Y. L. Kim, and Y. M. Song, "Revisiting silk: a lens-free optical physical unclonable function," *Nat. Commun.*, vol. 13, no. 1, p. 247, Jan. 2022, doi: 10.1038/s41467-021-27278-5.
[16] A. Anastasiou, E. I. Zacharaki, A. Tsakas, K. Moustakas, and D. Alexandropoulos, "Laser fabrication and evaluation of holographic intrinsic physical unclonable functions," *Sci. Rep.*, vol. 12, no. 1, p. 2891, Feb. 2022, doi: 10.1038/s41598-022-06407-0.
[17] C. Mesaritakis *et al.*, "Physical Unclonable Function based on a Multi-Mode Optical Waveguide," *Sci. Rep.*, vol. 8, no. 1, p. 9653, Jun. 2018, doi: 10.1038/s41598-018-28008-6.
[18] R. A. John *et al.*, "Halide perovskite memristors as flexible and reconfigurable physical unclonable functions," *Nat. Commun.*, vol. 12, no. 1, p. 3681, Jun. 2021, doi: 10.1038/s41467-021-24057-0.
[19] H. M. Ibrahim, H. Abunahla, B. Mohammad, and H. AlKhzaimi, "Memristor-based PUF for lightweight cryptographic randomness," *Sci. Rep.*, vol. 12, no. 1, p. 8633, May 2022, doi: 10.1038/s41598-022-11240-6.
[20] B. C. Grubel *et al.*, "Silicon photonic physical unclonable function," *Opt. Express*, vol. 25, no. 11, p. 12710, May 2017, doi: 10.1364/OE.25.012710.
[21] F. B. Tarik, A. Famili, Y. Lao, and J. D. Ryckman, "Scalable and CMOS compatible silicon photonic physical unclonable functions for supply chain assurance," *Sci. Rep.*, vol. 12, no. 1, p. 15653, Sep. 2022, doi: 10.1038/s41598-022-19796-z.
[22] A. M. Smith and H. S. Jacinto, "Reconfigurable Integrated Optical Interferometer Network-Based Physically Unclonable Function," *J. Light. Technol.*, vol. 38, no. 17, pp. 4599–4606, Sep. 2020, doi: 10.1109/JLT.2020.2996015.
[23] B. T. Bosworth *et al.*, "Unclonable photonic keys hardened against machine learning attacks," *APL Photonics*, vol. 5, no. 1, p. 010803, Jan. 2020, doi: 10.1063/1.5100178.
[24] D. Dermanis, A. Bogris, P. Rizomiliotis, and C. Mesaritakis, "Photonic Physical Unclonable Function Based on Integrated Neuromorphic Devices," *J. Light. Technol.*, vol. 40, no. 22, pp. 7333–7341, Nov. 2022, doi: 10.1109/JLT.2022.3200307.
[25] K. Sozos, A. Bogris, P. Bienstman, G. Sarantoglou, S. Deligiannidis, and C. Mesaritakis, "High-speed photonic neuromorphic computing using recurrent optical spectrum slicing neural networks," *Commun. Eng.*, vol. 1, no. 1, p. 24, Oct. 2022, doi: 10.1038/s44172-022-00024-5.

[26] A. Tsirigotis *et al.*, "Unconventional Integrated Photonic Accelerators for High-Throughput Convolutional Neural Networks," *Intell. Comput.*, vol. 2, p. 0032, Jan. 2023, doi: 10.34133/icomputing.0032.

[27] M. Lukoševičius, H. Jaeger, and B. Schrauwen, "Reservoir Computing Trends," *KI - Künstl. Intell.*, vol. 26, no. 4, pp. 365–371, Nov. 2012, doi: 10.1007/s13218-012-0204-5.

[28] A. F. Atiya and A. G. Parlos, "New results on recurrent network training: unifying the algorithms and accelerating convergence," *IEEE Trans. Neural Netw.*, vol. 11, no. 3, pp. 697–709, May 2000, doi: 10.1109/72.846741.

[29] M. C. Mackey and L. Glass, "Oscillation and Chaos in Physiological Control Systems," *Science*, vol. 197, no. 4300, pp. 287–289, Jul. 1977, doi: 10.1126/science.267326.

[30] L. E. Bassham *et al.*, "A statistical test suite for random and pseudorandom number generators for cryptographic applications," National Institute of Standards and Technology, Gaithersburg, MD, NIST SP 800-22r1a, 2010. doi: 10.6028/NIST.SP.800-22r1a.

[31] Y. Xing, J. Dong, S. Dwivedi, U. Khan, and W. Bogaerts, "Accurate extraction of fabricated geometry using optical measurement," *Photonics Res.*, vol. 6, no. 11, p. 1008, Nov. 2018, doi: 10.1364/PRJ.6.001008.

[32] Y. Su, Y. Zhang, C. Qiu, X. Guo, and L. Sun, "Silicon Photonic Platform for Passive Waveguide Devices: Materials, Fabrication, and Applications," *Adv. Mater. Technol.*, vol. 5, no. 8, p. 1901153, Aug. 2020, doi: 10.1002/admt.201901153.

[33] L. Devroye, *Non-Uniform Random Variate Generation*. New York, NY: Springer New York, 1986. doi: 10.1007/978-1-4613-8643-8.

[34] D. E. Knuth, *The art of computer programming, volume 2 (3rd ed.): seminumerical algorithms*. USA: Addison-Wesley Longman Publishing Co., Inc., 1997.